\documentclass[doublecol]{epl2} 

\usepackage{graphicx}
\usepackage{amsmath,amssymb}
\usepackage{bm}

\newcommand{\kB}{k_\mathrm{B}}
\newcommand{\bmf}{\bm{f}}
\newcommand{\bmxi}{\bm{\xi}}

\title{Energetics of single active diffusion trajectories}
\shorttitle{Energetics of single active diffusion trajectories}

\author{S. Shinkai\inst{1} \and Y. Togashi\inst{2}}
\shortauthor{S. Shinkai and Y. Togashi}

\institute{                    
  \inst{1} Department of Mathematical and Life Science, Graduate School of Science, Hiroshima University, Kagamiyama, Higashi-Hiroshima 739-8526, Japan\\
  \inst{2} Department of Computational Science, Graduate School of System Informatics, Kobe University, Rokkodai, Kobe 657-8501, Japan
}

\abstract{
The fundamental insight into Brownian motion by Einstein is that all substances exhibit continual fluctuations due to thermal agitation balancing with the frictional resistance.
However, even at thermal equilibrium,
biological activity can give rise to non-equilibrium fluctuations that cause ``active diffusion" in living cells.
Because of the non-stationary and non-equilibrium nature of such fluctuations,
mean square displacement analysis,
relevant only to a steady state ensemble,
may not be the most suitable choice as it depends on the choice of the ensemble;
hence, a new analytical method for describing active diffusion is desired.
Here we discuss the stochastic energetics of a thermally fluctuating single active diffusion trajectory driven by non-thermal random forces.
Heat dissipation,
usually difficult to measure,
can be estimated from the active diffusion trajectory;
guidelines on the analysis such as criteria for the time resolution and driving force intensity are shown by a statistical test.
This leads to the concept of an ``instantaneous diffusion coefficient" connected to heat dissipation that may be used to analyse the activity and molecular transport mechanisms of living systems.
}

\begin{document}

\maketitle

\section{Introduction}
Einstein's development of the theory of Brownian motion enabled advances such as the enumeration of Avogadro's number through mean square displacement (MSD) analysis of diffusive micro particles in equilibrium~\cite{Einstein1956, Perrin1913}
and has been generalized to the fluctuation-dissipation relation (FDR) in statistical physics~\cite{KuboTodaHashitsume1991}.

\begin{figure}[t]
	\onefigure[width=80mm]{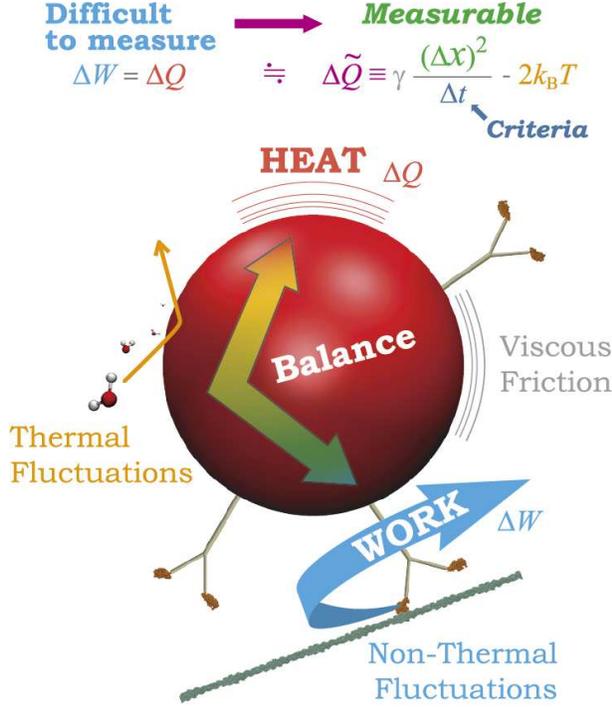}
	\caption{Schematic illustration of the energetics of an active diffusive particle.
	Under such fluctuations, dissipative heat $\Delta Q_n$ balances with the work $\Delta W_n$ done on the system~\cite{Sekimoto2010}.
	As it is usually experimentally difficult to measure the non-thermal force driving the particle in active diffusion, it is correspondingly nearly impossible to measure the work balance with the dissipative heat.
	However, the value $\Delta \widetilde{Q}_n = \gamma (\Delta x_n)^2 / \Delta t - 2 \kB T$,
	which can be estimated from the diffusion trajectory $\{ x_n \}$,
	can be regarded to be approximate dissipative heat.}
	\label{fig:1}
\end{figure}

However, many macromolecules in biological systems can exert their functions out of equilibrium by using external free energy from suppliers such as
ATP
molecules and dissipating heat to the environment;
such behaviour obviously violates the FDR and has been exposed by microrheology studies of both living cells~\cite{Caspi2000, Lau2003, Wilhelm2008} and in vitro cytoskeletal networks~\cite{Brangwynne2009, Mizuno2007}.
ATP-dependent non-equilibrium fluctuations violating the FDR cause active diffusion in living cells even at thermal equilibrium~\cite{Weber2012, MacKintosh2012}.
Understanding the energetics of non-equilibrium steady states enables the measurement of the extent of such violations in terms of the energy dissipation rate~\cite{Harada2005, HaradaSasa2005, HaradaSasa2007, Toyabe2010}
and the velocity FDRs~\cite{Speck2006}.
The activities of molecular machines such as motors can also affect the transport of molecules within a system.
Some molecular motors can carry cargoes such as vesicles or nanoparticles and move processively along a filament,
as occurs in the case of myosin working on an actin filament and kinesin on a microtubule.
Occasionally, a ``tug-of-war" between motors~\cite{Kural2005, Ali2011}, in which a single cargo is carried by two or more types of motors and changes either direction or tracks~\cite{Ross2008} (e.g. microtubule to actin), is also observed.
This situation can also be described as active diffusion, as cargoes are driven by both stochastically applied forces (non-thermal fluctuations applied by the motors) and thermal fluctuations (see fig.~\ref{fig:1}).

Recent single molecule measurements have provided insights into the mechanisms of molecular machines, both in vitro~\cite{Finer1994, Funatsu1995, Moerner1999, Weiss1999} and in vivo~\cite{Ueda2001, Sako2006}.
It remains difficult, however, to directly measure the force or work exerted by individual motors within a cell, especially when multiple motors work on single cargoes;
although the motion of a single molecule or nano-particle can be tracked, the number of samples that can be extracted is typically very limited and the motion measured may reflect the local cellular environment.
Thus,
when we routinely take an ensemble average over a set of trajectories,
most of the information on in vivo machine behaviour may be lost.
Thus, it would be useful to develop a methodology for extracting information from single trajectories; accordingly, we propose in this study a new method of measuring the heat dissipation of a single active diffusion trajectory.

\section{Active diffusion model}

We can use an overdamped Langevin system with a non-thermal and non-conservative random force $\hat{f}(t)$ to model active diffusion in living cells:
\begin{equation}\label{eq:1}
	\gamma \dot{x}(t) = \hat{f}(t) + \sqrt{2\gamma \kB T} \, \hat{\xi}(t),
\end{equation}
where $\gamma$ is the friction constant,
$\kB$ is the Boltzmann constant,
$T$ is the temperature of the environment,
and $\hat{\xi}(t)$ is Gaussian white noise satisfying $\langle \hat{\xi}(t) \hat{\xi}(s) \rangle = \delta(t-s)$.
$\langle \cdot \rangle$ denotes the ensemble average.
We assume that $\hat{f}(t)$ is an arbitrary stochastic process to express active driving force from the non-thermal environment and has a certain value in the time interval $[t, t+\delta t)$, in which the time resolution $\delta t$ makes the Langevin equation well-defined.
For simplicity, diffusion is assumed here to occur in one-dimensional space,
although it would be easy to extend the analysis to $d$-dimension ($d = 2$ or $3$).
More generally, a potential force $-dU(x)/dx$ should be added in eq.~(\ref{eq:1});
however, the Langevin equation can be effectively reduced to the same form as this equation for a periodic potential under certain conditions~\cite{HayashiSasa2005}.

At a time resolution $\Delta t$ and a discrete time $t_n \equiv n \Delta t$, an experimentally observed diffusion trajectory is representable by a discrete data set $\{x_n \equiv x(t_n)\}$.
Here, we introduce a Wiener process~\cite{Sekimoto2010, Gardiner2009}, $B(t)$,
instead of the noise $\hat{\xi}(t)$ to integrate eq.~(\ref{eq:1}).
Then, $\bmxi_n \equiv (B(t_{n+1}) - B(t_n)) / \sqrt{\Delta t}$ is the Gaussian random variable satisfying $\langle \bmxi_n \rangle = 0$ and $\langle \bmxi_n \bmxi_m \rangle = \delta_{nm}$ from the definition of the Wiener process.
The corresponding discrete Langevin equation can then be obtained by integrating eq.~(\ref{eq:1}) from $t_n$ to $t_{n+1}$:
\begin{equation}\label{eq:2}
	\gamma \Delta x_n = \bmf_n \Delta t+ \sqrt{2 \gamma \kB T \Delta t} \, \bmxi_n,
\end{equation}
where $\Delta x_n \equiv x_{n+1} - x_n$.
We assume that $\bmf_n$ is a random variable representing a random force in the time interval $[t_n, t_n+\Delta t)$.
Note that the difference between the approximate $x_n$ of eq.~(\ref{eq:2}) and the true $x(t)$ of eq.~(\ref{eq:1}) is $O(\Delta t^{1/2})$~\cite{Sekimoto2010},
and that numerical calculations are done by using of eq.~(\ref{eq:2}), called the Euler scheme~\cite{Gardiner2009},
because of the independence of $\bmf_n$ from the position $x_n$.

\section{Dissipative heat}

Thermodynamic variables such as work and heat in the fluctuating world
can be described using stochastic energetics~\cite{Sekimoto2010}.
The energy dissipated by a system to the environment during time interval $[t_n, t_{n+1})$ can be expressed as
\begin{equation}\label{eq:3}
	\Delta Q_n \equiv
	\left( \gamma \frac{\Delta x_n}{\Delta t} - \sqrt{\frac{2 \gamma \kB T}{\Delta t}} \bmxi_n \right)
	\circ \Delta x_n,
\end{equation}
where $\circ$ denotes Stratonovich multiplication.
This dissipated energy can be defined as the dissipative heat of the system, and because there is no potential force in eq.~(\ref{eq:1}),
this dissipative heat balances with the work done to the system by the random force $\bmf_n$ during the time interval $[t_n, t_{n+1})$:
\begin{equation}\label{eq:4}
	\Delta Q_n = \Delta W_n \equiv \bmf_n \circ \Delta x_n.
\end{equation}
As $\bmf_n$ is a non-conservative force (i.e. it does not depend on $x_n$),
Stratonovich multiplication $\circ$ is equivalent to usual multiplication in all of the formulation that follows.
Combining eqs.~(\ref{eq:2})-(\ref{eq:4}),
we obtain the energy balance relation:
\begin{equation}\label{eq:5}
	\Delta W_n = \Delta Q_n = \Delta \widetilde{Q}_n + 2 \kB T (1 - \bmxi_n^2)
	- \sqrt{\frac{2 \kB T \Delta t}{\gamma}} \bmf_n \bmxi_n,
\end{equation}
where
\begin{equation}\label{eq:6}
	\Delta \widetilde{Q}_n \equiv \frac{\gamma (\Delta x_n)^2}{\Delta t} - 2\kB T.
\end{equation}
Note that the work and the dissipative heat can be determined only when all of the information on $x_n$ and $\bmf_n$ are known;
however, as the random force $\bmf_n$ cannot usually be observed,
we can instead estimate the quantity $\Delta \widetilde{Q}_n$ using only the observed diffusion trajectory $\{ x_n \}$.
Equation~(\ref{eq:6}) implies that $\Delta \widetilde{Q}_n$ equals the excess energy left over in subtracting the average energy dissipation in equilibrium $2 \kB T$ from the energy generated by the motion $\gamma (\Delta x_n)^2 / \Delta t$.
Here $\Delta Q_n \neq \Delta \widetilde{Q}_n$;
nevertheless, given the ergodicity of $\bmxi_n$,
$\displaystyle \langle \bmxi_n^2 \rangle = \lim_{N \to \infty} \frac{1}{N} \sum_{n=0}^{N-1} \bmxi_n^2 = 1$,
and the independence between the random variables $\{ \bmxi_n \}$ and $\{ \bmf_n \}$, we see that
$\displaystyle \lim_{N \to \infty} \frac{1}{N} \sum_{n=0}^{N-1} \bmf_n \bmxi_n = \left( \lim_{N \to \infty} \frac{1}{N} \sum_{n=0}^{N-1} \bmf_n \right) \langle \bmxi_n \rangle = 0$, and thus
the energy dissipation rates calculated from the long-time averages of both $\Delta Q_n$ and $\Delta \widetilde{Q}_n$ are equivalent:
$\displaystyle \lim_{N \to \infty} \frac{1}{N\Delta t} \sum_{n=0}^{N-1} \Delta Q_n = \lim_{N \to \infty} \frac{1}{N\Delta t} \sum_{n=0}^{N-1} \Delta \widetilde{Q}_n$.

\begin{figure}[t]
	\onefigure[width=80mm]{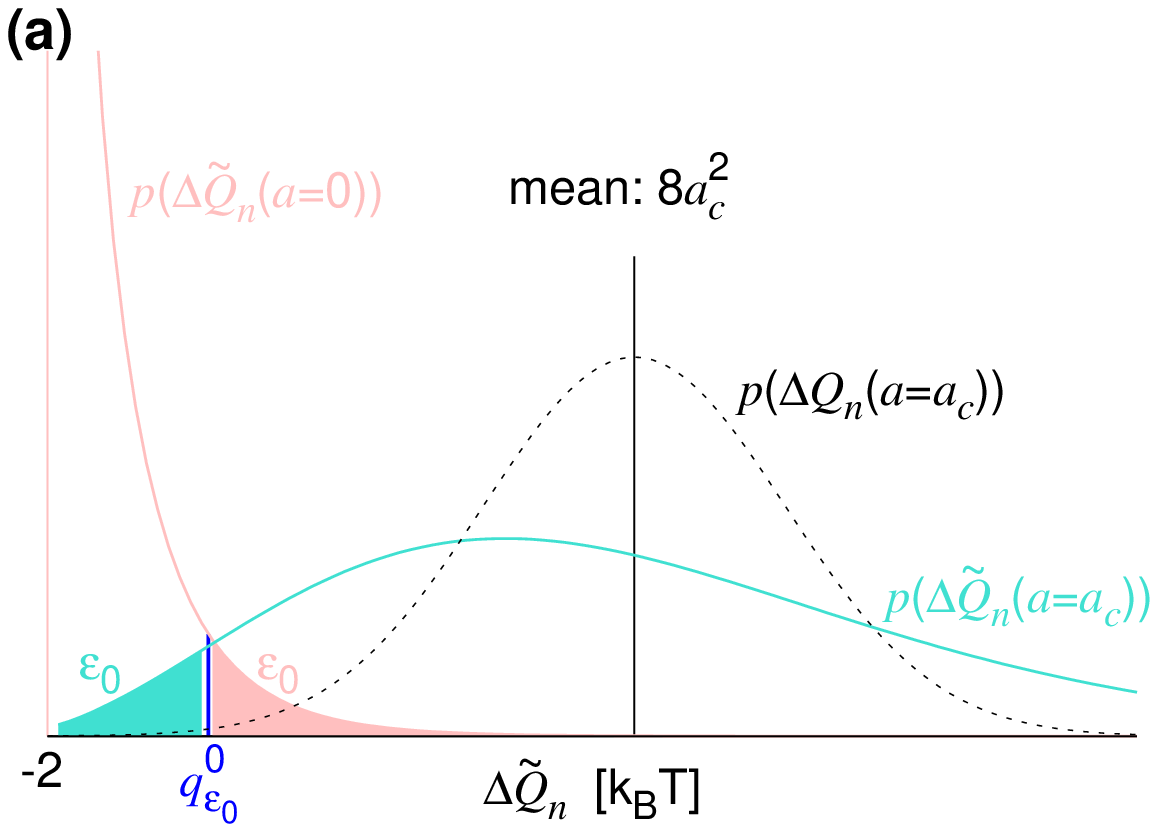}
	\onefigure[width=80mm]{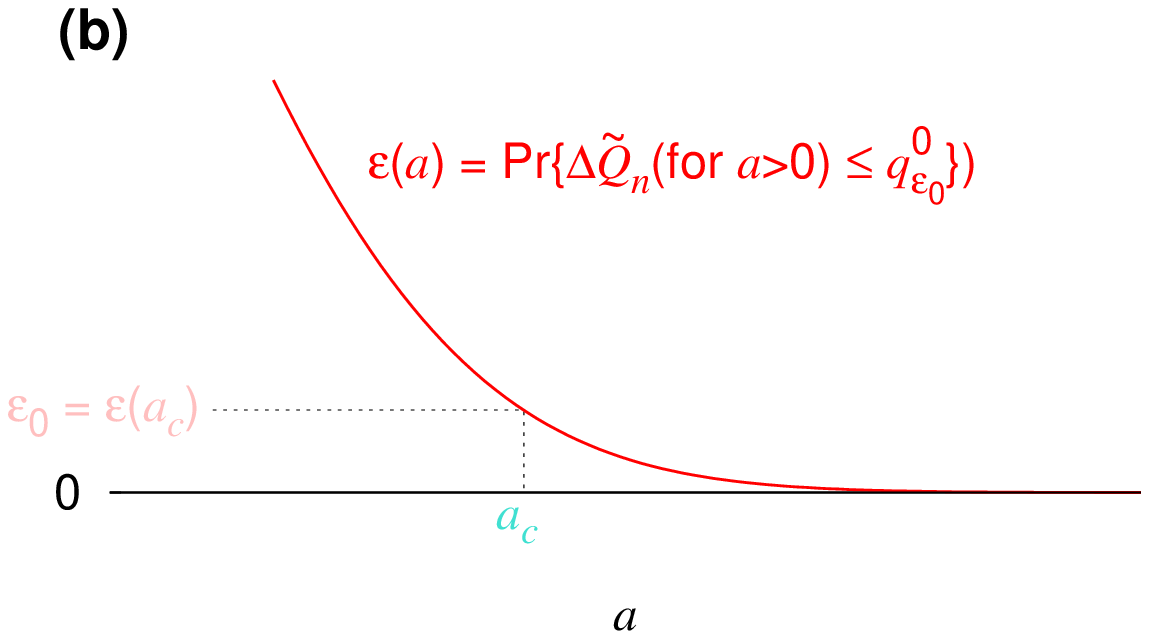}
	\caption{Statistical test of $\Delta \widetilde{Q}_n$.
	(a)
	Distributions of $\Delta \widetilde{Q}_n$ for $a=0$, $a = a_c$, and $\Delta Q_n$ for $a=a_c$.
	For a fixed significance level $\varepsilon_0$, the relation among the filled regions and $q_{\varepsilon_0}^0$ is: $\Pr\{ \Delta \widetilde{Q}_n(a=0) > q_{\varepsilon_0}^0 \} = \Pr\{ \Delta \widetilde{Q}_n(a=a_c) \le q_{\varepsilon_0}^0 \} = \varepsilon_0$.
	$\Delta Q_n$ and $\Delta \widetilde{Q}_n$ have the same mean value $8a^2$.
	(b)
	The function $\varepsilon(a)$, which corresponds to the probability that $\Delta \widetilde{Q}_n$ for $a>0$ is less than $q_{\varepsilon_0}^0$,
	is a decreasing function of $a$ (see Appendix).
	The value $a_c$ satisfies $\varepsilon(a_c) = \varepsilon_0$.}
	\label{fig:2}
\end{figure}

\section{Statistical test}
As shown in eqs.~(\ref{eq:5}) and (\ref{eq:6}), the energy dissipation values
$\Delta Q_n$ and $\Delta \widetilde{Q}_n$ are fluctuating quantities.
To better understand their distributions,
we can use the following non-dimensional parameter for a random force $\bmf_n = f$ applied to the system at $t_n$:
\begin{equation*}
	a = \sqrt{\frac{f^2 \Delta t}{8 \gamma \kB T}}.
\end{equation*}
Using the random variable $\bmxi_n$, the dissipation values can be written as
\begin{equation*}
	\Delta Q_n = 4a (\bmxi_n + 2a) \quad \mbox{and} \quad
	\Delta \widetilde{Q}_n = 2 (\bmxi_n + 2a)^2 - 2
\end{equation*}
in $\kB T$ energy units.
Because they have the same mean value $8a^2$, $\langle \Delta Q_n \rangle = \langle \Delta \widetilde{Q}_n \rangle = 8a^2$ for any $a \ge 0$,
and thus both share the same average behaviour.
However,
$\Delta \widetilde{Q}_n$ has larger fluctuations owing to its larger standard deviation $\Delta \widetilde{Q}_n$ = $8 \sqrt{a^2 + 1/8}$ , as compared to $4a$ for $\Delta Q_n$.
Moreover, the relative fluctuation of this difference decreases as a function of $a$:
$\displaystyle \frac{\sqrt{\langle (\Delta \widetilde{Q}_n - \Delta Q_n)^2 \rangle}}{\langle \Delta Q_n \rangle} = \frac{1}{2a} \sqrt{1 + \frac{1}{2a^2}}$.

As mentioned above,
the value $\Delta \widetilde{Q}_n$, but not $\Delta Q_n$, can be evaluated from observations of a diffusion trajectory $\{ x_n \}$; correspondingly,
our primary result is that we can statistically distinguish the distribution of $\Delta \widetilde{Q}_n$ for $a > 0$,
where the system is driven by an active random force,
from that for the equilibrium condition $a = 0$.
This is derived from the following statistical consideration:
The pink, green and dotted curves in fig.~\ref{fig:2}(a) show the probability densities of $\Delta \widetilde{Q}_n$ for $a=0$, $\Delta \widetilde{Q}_n$ for $a = a_c$ and $\Delta Q_n$ for $a=a_c$, respectively.
Since the distribution of $\Delta \widetilde{Q}_n$ for $a=0$ is uniquely determined,
for a fixed significance level $\varepsilon_0$,
the value $q_{\varepsilon_0}^0$ satisfying $\Pr \{ \Delta \widetilde{Q}_n (a=0) > q_{\varepsilon_0}^0 \} = \varepsilon_0$ is also uniquely determined.
This probability corresponds to the pink filled region in fig.~\ref{fig:2}(a).
For these $\varepsilon_0$ and $q_{\varepsilon_0}^0$,
let us define a function $\varepsilon(a)$ as the probability $\Pr \{ \Delta \widetilde{Q}_n (\mbox{for $a>0$}) \le q_{\varepsilon_0}^0 \}$.
Then, $\varepsilon(a)$ is explicitly written as eq.~(\ref{eq:a2}) (see Appendix),
and is a decreasing function of $a$ (see fig.~\ref{fig:2}(b)).
Therefore there exists a unique $a_c$ such that $\varepsilon(a_c) = \Pr\{ \Delta \widetilde{Q}_n(a=a_c) \le q_{\varepsilon_0}^0 \} = \varepsilon_0$,
and the inequality $q_{\varepsilon_0}^0 < q_{\varepsilon_0}(a)$ holds for $a > a_c$,
where the value $q_{\varepsilon_0}(a)$ satisfies $\Pr \{ \Delta \widetilde{Q}_n (a>0) \le q_{\varepsilon_0}(a) \} = \varepsilon_0$.
In other words,
the upper confidence bound $q_{\varepsilon_0}^0$ of $\Delta \widetilde{Q}_n$ for $a = 0$ is smaller than the lower bound $q_{\varepsilon_0}(a)$ for $a > a_c$.
The green filled region, which stands for the probability $\Pr\{ \Delta \widetilde{Q}_n(a=a_c) \le q_{\varepsilon_0}^0 \}$,
represents the false negative rate at which active diffusion ($a = a_c$) is misjudged to be in equilibrium ($a=0$),
while the pink filled region denotes the false positive rate at which equilibrium fluctuations are misjudged to be active fluctuations in fig.~\ref{fig:2}(a).
For example, this numerical calculation results in $a_c = 1.803$ at the significance level $\varepsilon_0 = 0.05$ (see Table \ref{ta:1}).

The above statistical consideration
provides criteria for the time resolution and driving force,
namely:
\begin{equation*}
	\Delta t > \frac{8 \gamma \kB T}{f^2} a_c^2 \quad \mbox{and} \quad
	|f| > \sqrt{\frac{8 \gamma \kB T}{\Delta t}} a_c,
\end{equation*}
respectively.
The former implies the existence of the lower bound of $\Delta t$ with which one can statistically distinguish active fluctuations from thermal equilibrium fluctuations for a known driving force,
and the latter vice versa.

\begin{figure*}[t]
	\onefigure[width=165mm]{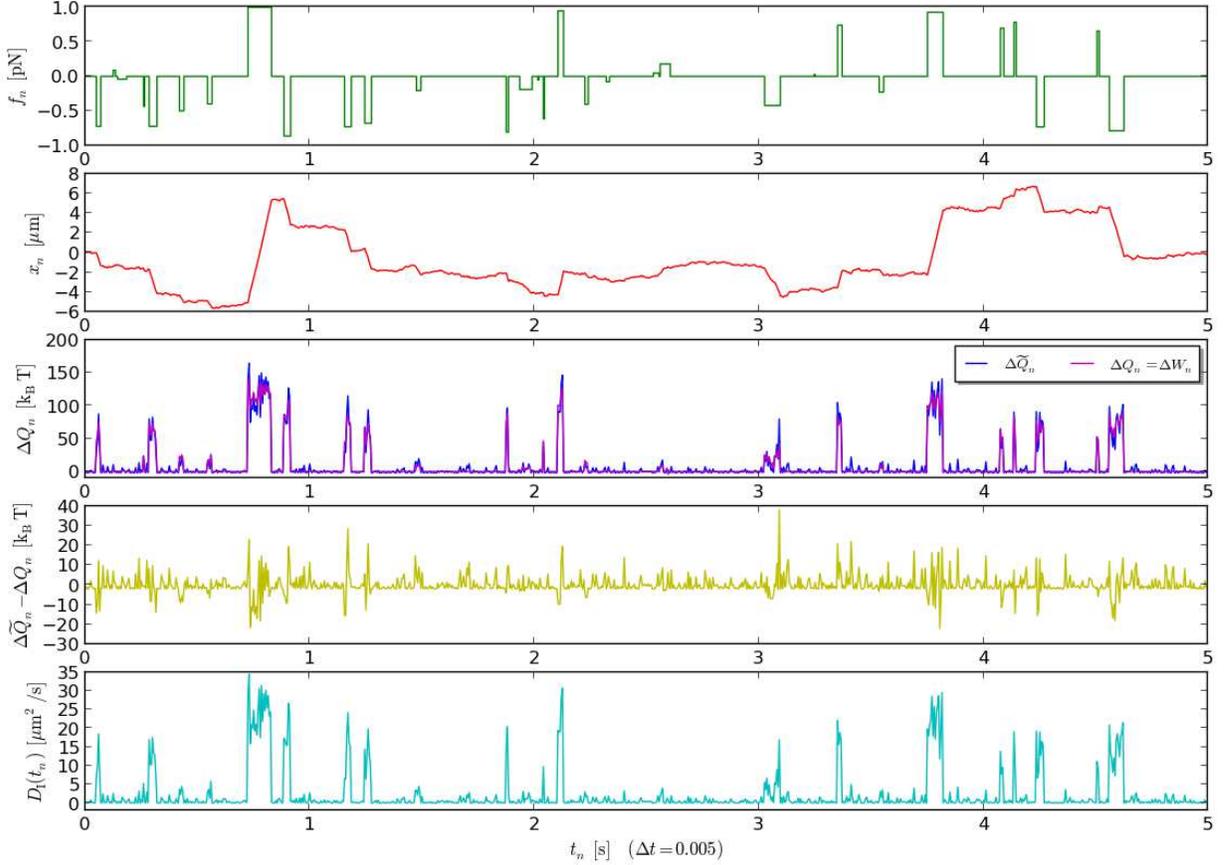}
	\caption{
	Numerical trajectories of external random force, active diffusion, dissipative heats, the difference between $\Delta \widetilde{Q}_n$ and $\Delta Q_n$, and instantaneous diffusion coefficient
	for the parameters $\gamma = 10^{-8} \un{kg/s}$, $T=300 \un{K}$ and $\Delta t = 5 \un{ms}$.
	The equilibrium diffusion coefficient becomes $D_\mathrm{eq} = \kB T / \gamma \simeq 0.414 \un{\mu m^2/s}$.
	}
	\label{fig:3}
\end{figure*}

\section{Numerical demonstration}
Figure~\ref{fig:3} shows numerical trajectories for a random force $\{ \bmf_n \}$ from which a diffusion trajectory $\{ x_n \}$ is obtained using eq.~(\ref{eq:2}).
Given external random forces $\{ \bmf_n \}$, the trajectory $\{ x_n \}$ is calculated by using the integral Langevin equation~(\ref{eq:2}) for $\gamma = 10^{-8} \un{kg/s}$, $T=300 \un{K}$ and $\Delta t = 5 \un{ms}$.
Behaviours of both $\Delta \widetilde{Q}_n$ and $\Delta Q_n$ are almost same with the fluctuating difference $\Delta \widetilde{Q}_n - \Delta Q_n = 2(\bmxi_n^2 - 1) + 8 a \bmxi_n$,
where the value $a$ is also a random variable $\sqrt{\bm{f}_n^2 \Delta t / (8 \gamma k_\mathrm{B} T)}$ and the mean value $\langle \Delta \widetilde{Q}_n - \Delta Q_n \rangle$ becomes zero.
At a significance level of $\varepsilon_0 = 0.05$, the condition $a > a_c$ for distinguishing the distributions between $\Delta \widetilde{Q}_n$ for $a=0$ and $a>0$ can be transformed into $|\bmf_n| \gtrsim 0.424 \un{pN}$.
For a system driven by random force satisfying this condition,
$\Delta \widetilde{Q}_n$ can be distinguished from the dissipative heat generated by thermal fluctuations.

\section{Discussion}
A physical interpretation of the value of $\{ \Delta \widetilde{Q}_n \}$ is that it represents the measurable dissipative heat trajectory of a diffusion trajectory $\{ x_n \}$.
The formulation of $\Delta \widetilde{Q}_n$ developed here represents an alternative to MSD analysis.
If we rewrite eq.~(\ref{eq:6}) as
$\displaystyle \Delta \widetilde{Q}_n
= 2 \gamma \left[ \frac{(\Delta x_n)^2}{2 \Delta t} - \frac{\kB T}{\gamma} \right]
= 2 \gamma \left[  D_\mathrm{I}(t_n) - D_\mathrm{eq} \right]$,
then, because of the Einstein relation~\cite{KuboTodaHashitsume1991},
in which $\kB T / \gamma$ is equivalent to the diffusion coefficient $D_\mathrm{eq}$ for the thermal equilibrium Langevin equation~(\ref{eq:1}) for $\hat{f}(t) = 0$,
the instantaneous diffusion coefficient:
\begin{equation*}
	D_\mathrm{I}(t_n) \equiv \frac{(\Delta x_n)^2}{2 \Delta t}
\end{equation*}
will be an energetically well-defined physical quantity.
Although this definition is trivial in the context of stationary diffusion described by the diffusion equation~\cite{Einstein1956},
in which the relation $D_\mathrm{eq} = \langle D_\mathrm{I}(t_n) \rangle$ is satisfied,
it is noteworthy that the instantaneous diffusion coefficient developed here captures the essential characteristics of active diffusion even within a non-stationary active environment at an appropriate time resolution $\Delta t$:
The instantaneous diffusion coefficient is agitated when the system is driven by active force;
in thermal equilibrium, on the other hand,
it fluctuates around the equilibrium diffusion coefficient $D_\mathrm{eq}$;
the calculations of $D_\mathrm{I}(t_n)$ are independent of the ensemble of diffusion trajectories.
Because of this,
we can use the instantaneous diffusion coefficient $D_\mathrm{I}(t_n)$ to investigate single active diffusion trajectories such as those occurring in the molecular transport mechanisms of living systems:
When the instantaneous diffusion coefficient drifts upward from that for thermal fluctuations,
the probe particle would be driven by certain active force;
for example,
in a non-equilibrium gel consisting of myosin II, actin filaments, and cross-linkers~\cite{Mizuno2007},
the source of the active force is motor activity using ATP.

Note that diffusion ``trajectories" in this paper does not correspond to the definition of ``state" in the Jarzynski and Crooks non-equilibrium work relations~\cite{Jarzynski1997, Crooks1998, Esposito2010}.

In this study, we demonstrated that it is possible to extract physical quantities such as dissipative heat and the related instantaneous diffusion coefficient from individual active diffusion trajectories.
Our approach differs from that
of the pioneering works in non-equilibrium steady state energetics~\cite{Harada2005, HaradaSasa2005, HaradaSasa2007, Toyabe2010} for measuring mechanical system responses;
based on our work,
it is now possible to estimate dissipative heat,
without measuring mechanical responses,
simply by tracking a single diffusion trajectory.
As living cells change their internal environments in complicated ways
with consuming energy and dissipating heat
throughout the cell cycle and in response to the external environment,
it will be interesting to use the instantaneous diffusion coefficient to quantify these adaptive changes.
In addition, by using this method it should be possible to determine the number of active motors attached to individual cargoes and to analyse their methods of cooperation.

\setcounter{table}{0}
\renewcommand{\thetable}{A\arabic{table}}
\section{Appendix: Detailed estimation of statistical distinction between dissipative heat under thermal equilibrium and active fluctuation}

Because the random variable $\bmxi_n$ has a Gaussian distribution with mean $0$ and variance $1$,
\begin{equation*}
	G(x) = \Pr \{ \bmxi_n \le x \} = \int_{-\infty}^x g(y) \, \upd y,
\end{equation*}
where
\begin{equation*}
	g(y) = \frac{1}{\sqrt{2\pi}} \, e^{-y^2/2},
\end{equation*}
the distribution of $\Delta \widetilde{Q}_n$ can then be calculated as follows:
\begin{eqnarray*}
	\Pr\{ \Delta \widetilde{Q}_n \le x \} & = &
	\Pr\{ 2(\bmxi_n +2a)^2 - 2 \le x \}\\
	& = & \Pr \left\{ - \sqrt{\frac{x}{2} + 1} \le \bmxi_n+2a \le \sqrt{\frac{x}{2} + 1} \right\}\\
	& = & G(h(x) - 2a) - G(-h(x)-2a),
\end{eqnarray*}
where
\begin{equation*}
	h(x) = \sqrt{\frac{x}{2} + 1}.
\end{equation*}
Here, we define the statistical distinction of $\Delta \widetilde{Q}_n$ for $a=0$ and $a>0$ as the following inequality at a significance level $0 < \forall \varepsilon_0 < 1$:
\begin{equation}\label{eq:a1}
	q_{\varepsilon_0}^0 < q_{\varepsilon_0}(a), \tag{A.1}
\end{equation}
where the values $q_{\varepsilon_0}^0$ and $q_{\varepsilon_0}(a)$ satisfy
\begin{equation*}
	\Pr \{ \Delta \widetilde{Q}_n (a=0) \le q_{\varepsilon_0}^0 \} =
	G(h(q_{\varepsilon_0}^0)) - G(-h(q_{\varepsilon_0}^0)) = 1 - \varepsilon_0
\end{equation*}
and
\begin{eqnarray*}
	& & \Pr \{ \Delta \widetilde{Q}_n (a>0) \le q_{\varepsilon_0}(a) \}\\
	& & = G(h(q_{\varepsilon_0}(a)) -2a) - G(-h(q_{\varepsilon_0}(a)) -2a) = \varepsilon_0,
\end{eqnarray*}
respectively.
Note that the value $q_{\varepsilon_0}^0$ is a constant for a fixed $\varepsilon_0$.
The probability that $\Delta \widetilde{Q}_n$ for $a>0$ is less than $q_{\varepsilon_0}^0$ depends on $a$ and can be expressed as
\begin{equation}\label{eq:a2}
	\varepsilon(a) = G(-2a+h_0) - G(-2a-h_0), \tag{A.2}
\end{equation}
where $h_0 = h(q_{\varepsilon_0}^0)$.
We can then show that the probability $\varepsilon(a)$ monotonically decreases for $a>0$, that is,
the derivation $\varepsilon'(a)$ is negative, as follows:
\begin{eqnarray*}
	\frac{\upd \varepsilon(a)}{\upd a} & = & -2\left\{ g(-2a+h_0) - g(-2a-h_0) \right\} \\
	& = & - \sqrt{\frac{8}{\pi}} \, e^{-(4a^2+h_0^2)/2} \sinh(2h_0 a) < 0
\end{eqnarray*}
for $a>0$.
This inequality implies that there exists a unique $a_c > 0$ such that $\varepsilon(a_c) = \varepsilon_0$ and $q_{\varepsilon_0}^0 = q_{\varepsilon_0}(a_c)$ for a fixed $0 < \forall \varepsilon_0 < 1$,
and therefore the inequality (\ref{eq:a1}) holds for $a > a_c$.
 Based on this, we can statistically distinguish between the heat $\Delta \widetilde{Q}_n$ of a state in thermal equilibrium ($a=0$) from that in a state driven by active fluctuations for $a>a_c$.
Table \ref{ta:1} shows numerically calculated values of $a_c$ and $q_{\varepsilon_0}^0$ for some significance levels $\varepsilon_0$.

\begin{table}[t]
	\caption{Numerically calculated values of $a_c$ and $q_{\varepsilon_0}^0$ for some significance levels $\varepsilon_0$.}
	\label{ta:1}
	\begin{center}
		\begin{tabular}{ccc}
			$\varepsilon_0$ & $q_{\varepsilon_0}^0$ $[\kB T]$ & $a_c$\\
			0.10	& 3.4110 & 1.463\\
			0.05 & 5.6829 & 1.803\\
			0.01 & 11.2697 & 2.451
		\end{tabular}
	\end{center}
\end{table}

\acknowledgments
This work was supported by MEXT, Japan (KAKENHI 23115007).

\bibliographystyle{eplbib}

\end{document}